# The Essence of the Essence from the Web: The Metasearch Engine

[1]Rajender Nath, [2] Satinder Bal

[1]Kurukshetra University, Haryana, rnath_2k3@rediffmail.com
[2]Vaish College of Engineering.Haryana., :satinder_bal@yahoo.com

*Abstract: The exponential growth of information source on the web and in turn continuing technological progress of searching the information by using tools like Search Engines gives rise to many problems for the user to know which tool is best for their query and which tool is not. At this time Metasearch Engine comes into play by reducing the user burden by dispatching queries to multiple search engines in parallel and refining the results of these search engines to give the best out of best by doing superior job on their side. These engines do not own a database of Web pages rather they send search terms to the databases maintained by the search engine companies, get back results from all the search engines queried and then compile the results to be presented to the user. In this paper, we describe the working of a typical metasearch engine and then present a comparative study of traditional search engines and metasearch engines on the basis of different parameters and show how metasearch engines are better than the other search engines.*

Keywords: Metasearch Engine, Search Engine, Query, *Web, database*

## 1. INTRODUCTION

Today the World-Wide Web has become so popular that the Internet now is one of the popular means of publishing information [1]. But how to locate data in the ocean of information is essential and an urgent problem [2]. To simplify the problem of getting relevant results based on their search query, the Internet search engines were created that allowed getting lot of information from the world-wide-web in the form of web pages [1].

Search Engines are among the most successful application on the web today. They act as a system for searching the information available on the web by automatically searching the contents of other systems and creating a database of the results [3]. The most famous search engines include AltaVista, Infoseek, Google, and MSN. They provide good searching ability by indexing more pages on the web and maintaining the updated indices in their databases. Excluding the above mentioned search engines which are general search engines there are likely thousands of specialty search engines, for example, FirstGov that provides search from US Government, US public and US state, AskJeeves is a search service that aim to direct you to the exact page answering your query [2].

Despite so many search engines are available to help users in finding the information of their interest, searching on the web is not an easy task. The Problem is due to the vast amount of data on the web and its rapid updation and growth [2].

A recent survey [4] indicates that the web has approximately 550 billion web pages and only 1% of them are on the surface of the web while the rest are in the deep web. The coverage of each search engine is limited. Even the largest and strongest database can index only 5 billion pages. This causes information search to be difficult and irrelevant and some times incomplete information from a single search engine. The survey also shows that only 1.1% of first page results of different search engines are identical. It points to the importance of searching the web with multiple engines to find more top ranked results. Obviously it will bring inconvenience and difficulty to the user [5] [6].

This provides basis to the necessity of increasing search coverage via combining results of multiple search engines [4]. Metasearch Engines are designed to address these problems [2]. Metasearch Engines are powerful tools that send user query simultaneously to several search engines, web directories and sometimes to deep web and their databases of web pages, within a few seconds, you get results back from all the search engines queried [7]. Metasearch Engines do not compile a physical database or catalogue of the web pages. Instead, they take a user's request, pass it to several other heterogeneous databases and then compile the results in a homogeneous manner based on a specific algorithm and create virtual database [8]. Many Metasearch Engines are available now such as Metacrawler, Savvysearch, Cyber, Dogpile, Profusion [2]. No two Metasearch engines are alike. Some search only the most popular search engines while others also search lesser-known engines, newsgroups, and other databases. They also differ in how the results are presented and the number of engines that are used. Some metasearch engines list results according to search engine or database. Others return results according to relevance, often concealing which search engine returned which results. This benefits the user by eliminating duplicate hits and grouping the most relevant ones at the top of the list [8].

## 2. RELATED WORK

The growth of Internet is unparalleled by any previous developments. This gave origin to search engine in 1990. The first search engine created was Archie by Alan Emtage, a student at McGill University in Monteral [10].

In 1992, Veronica appeared on the scene. Soon after another user interface name Jughead appeared with the same purpose as veronica in 1993, both of these were used for files to be sent via Gopher [11] [12].

At this time, there was no World Wide Web. The main way people shared data was via File Transfer Protocol (FTP). In the meantime, Tim Berner Lee created World Wide Web. By 1993, the Web was beginning to change. Rather than being populated mainly by FTP sites, Gopher sites, and e-mail servers, web sites began to proliferate. In response to this change, Matthew Gray introduced World Wide Web Wanderer. ALIWEB was developed as the web page equivalent to Archie and Veronica.

The next development in cataloging the web came late in 1993 with spiders. Like robots, spiders scoured the web for web page information. These early versions looked at the titles of the web pages, the header information, and the URL as a source for key words [10].

The first popular search engine, Excite, has it roots in these early days of web cataloging. It was released for general use in 1994. The first full-text search engine was WebCrawler. Again in 1994, Yahoo! was developed and became the first popular searchable directory.

The next search engine to appear on the web was Lycos. The next major player in the search engine wars was Infoseek. The Infoseek search engine itself was unremarkable and showed little innovation beyond WebCrawler and Lycos. What made Infoseek stand out was its deal with Netscape to become the browser's default search engine replacing Yahoo!.

By 1995, Digital Equipment Corporation (DEC) introduced AltaVista. This search engine contained some innovations that set it apart from the others. First, it ran on a group of DEC Alpha-based computers. At the time, these were among the most powerful processors in existence. It was also the first to implement the use of Boolean operators (and, or, but, not) to help refine searches. Then HotBot appeared and became the most powerful search engine, by indexing more than 10,000,000 pages a day.

In 1995, a new type of search engine was introduced - The Metasearch Engine. The concept was simple. It would get key words from the user either by the user typing key words or a question and then forward the keywords to all of the major search engines. These search engines would send the results back to the metasearch engine and the metasearch engine would format the hits all on one page for concise viewing.

The first of these search engines was Metacrawler. There are also other major metasearch engines like ProFusion, Dogpile, Ask Jeeves, and C-Net's Search.com. Ask Jeeves combines many of the features such as natural language queries with the ability to search using several different search engines. C-Net's entry claims to use over 700 different search engines to obtain its results [10].

## 3. ARCHITECTURE OF METASEARCH ENGINE

Searching distributed databases requires three problems to be solved. They include 1) identification and characterization of primary search services whose results are to be merged; 2) selection of a subset of available search engines efficiency or effectiveness; 3) translation of the searcher's request into the relevant query language of each primary search service and getting/parsing results. Metasearch Engines can conceptually be dissected into a number of cooperating software components to overcome these problems Figure 1 adapted from [9] shows a typical architecture of a metasearch Engine. In the following, we briefly outline the functionality served by each of the components in the Metasearch process.

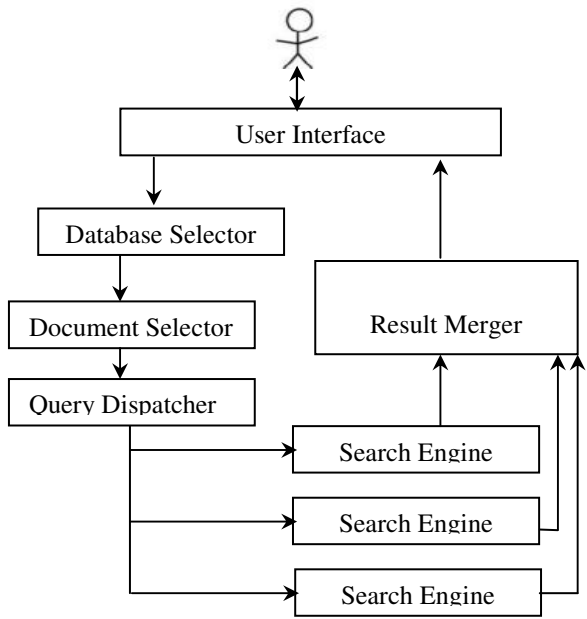

**Figure 1**. TYPICAL ARCHITECTURE OF A METASEARCH ENGINE

*Database Selector:* It is a software component that decides which group of search engines will be asked to search a given user query. The goal is to identify as many potentially useful sources as possible while minimizing the inclusion of useless ones.

The existing solutions to the database selection problem have been categorized into the following approaches: Rough representative, Statistical representative, and Learning-based [9]. The problem of selecting the information sources to be used was considered in a broader scope by including both search engines and databases. Hence the term Database Selection is the most commonly used term [9].

The main criterion used for determination of inclusion of web search engines is the availability of some API (Application Program Interface) to access that search service. Web service providers provide Search APIs to enable application developers to leverage the dynamic information generation capability afforded by search engines in the applications that they develop. Metasearch Engines will not leave any of the search engines, it is on the user to choose or leave search engine according to their preference and search query.

*Document Selector:* This component determines which results shall be retrieved from each of the databases so as to maximize useful documents while minimizing unrelated documents and network traffic. The criteria used for document selection can be specified by the user, based on statistical-based information, or learning-based information.

Document selection, is concerned with the issue of determining how many documents are to be selected and how is the quota distributed among the component search engines. The types of issues that arise in this process are commonly caused by the existence of a large number of component engines to choose from, and that some of them work with more focused datasets [4].

*Query Dispatcher:* It establishes connection with each individual search service, forward the query string, and wait to collect the results. This can be implemented by HTTP protocol using GET and POST methods, or via other web service protocols such as SOAP or REST. The dispatcher connects the metasearch engine to the search engines.

*Result Merger:* This component merges the result from different search engines by applying its merging algorithm and then presents the result to the user. It presents the result in two ways. One way is to simply list ten or more results from each engine queried with no additional post-processing. Dogpile works this way.

Other metasearch engines analyze the results and then rank them according to their own rules, combining results from multiple engines into a single, unified list. IxQuick, Metacrawler and Vivisimo are examples of this type of result aggregating metasearch engine. When results are retrieved separately from each of the search engines, a single ranked list needs to be assembled for presentation. This is the task of the result merger. The main complexity associated with the merging process is determining how to re-rank the results in the presence and absence of their ranking scores assigned by their respective search engines viewing. Each individual component search engine will return results ranked according to their own ranking algorithm. The only information that can be derived is the relative ordering among the documents in each result.

Metasearch Engine neither knows the exact ranking algorithm used, nor the actual scores associated with each document. Therefore, merging algorithm must be designed to work without either piece of information. Even if the scores are available somehow, the result merger is still faced with the issue of comparability among the scores. Duplicate results also contribute to the complexity of ranking. Some or all of the component search engines in response to a query may return a document. Result merging is concerned with presenting a single consistent result set to the end-user in response to a query. The design of the ranking algorithm is complicated by the fact that a duplicated result can be ordered differently in different search engines. It requires reconciling two or more rankings of the same result [9].

## 4. COMPARATIVE STUDY

After having created sufficient foundation, now, we compare the traditional search engines and metasearch engines on various parameters. The parameters that have been identified for comparative study are shown in Table 1. These parameters act as a measure of effectiveness of any information retrieval system or search service providers. In the comparative study, we derive how each single parameter is crucial in determining the effectiveness of both search engines and metasearch engines.

By critically analyzing the working of traditional search engines and Metasearch engines, we have compared these two categories of search engines on the parameters shown in Table 1. The results of comparative study have been summarized in Table 2. In the following, the results are discussed point wise.

*Size of database:* Search Engines need huge amount of database to enhance their relative performance and for providing more accurate information but unlike search engines, metasearch engines do not need to have their own database or index of web pages instead they often create a virtual database and simultaneously searches several databases. It gives metasearch engines an additional advantage of requiring little storage over search engines and result in more economical search service than search engine.

*Coverage:* Metasearch Engines offer the potential to search a larger portion of the Web than a single search engine, as search engines are able to perform their search only on the surface of the web and to some extent in the deep web. Thus, metasearch engines rather than search engines may become essential.

*Precision:* It can be defined as the ratio of retrieved relevant documents to the number of retrieved documents. Metasearch engines provide more precise results than traditional search engines by using query modification.

*Result Relevancy:* Any search service would prove to be beneficial to the user only if it is able to provide relevant and to the point information to the user. Thus, result relevancy can be considered as the most dominating criteria for any information retrieval system. Metasearch Engines supersede the traditional search engines in providing better scope and relevant and précised response by clustering and merging the top ranked results of various search engines.

*Query Response Time and Relevant Response Time:* Query response time is the time taken by any search service in providing the response to the user so as to equip the user with desired information. It is one of the major factors for assessing the performance of any search service. Although search engines provide the user with response within a blink of eye but these responses may not always be relevant and contain redundant information. Thus, for relevant results, user has to query different search engines, which increase the overall relevant response time for the user. Despite of querying different search engines, querying a single metasearch engine saves time of the user by providing relevant results.

There is tradeoff between the above two mentioned parameters, enhancing the one can have negative effect on the other.

*Network Bandwidth:* The search engines use crawlers or spiders to crawl the pages in the World Wide Web to update their databases continuously. This consumes a major portion of network bandwidth. Metasearch engines do not have database as they create a virtual database from the results of other search engines.

*Dependency:* For providing results search engines have to depend upon their ranking algorithm and their database while metasearch engines depend upon the search engines to be used.

*Redundancy:* The presence of duplicated links/documents in the result degrades the quality of information retrieved. For better quality result search engine depend on their ranking algorithm while, merging algorithm help metasearch engines in integrating the results in a better way, by eliminating duplicate results.

*Hardware Requirement and Implementation Cost:* Search Engines require large hardware for storage of database. Thus, increasing initial infrastructure and implementation cost. Storage requirement of metasearch engine is very less than search engine resulting in much smaller investment in hardware.

| Parameters |
| --- |
| Database size |
| Coverage |
| Precision |
| Result Relevancy |
| Response Time |
| Relevant result Time |
| Network Bandwidth |
| Dependency |
| Redundancy |
| Hardware Requirement |
| Implementation cost |

Table 1 Parameter List

| Search Engine Parameters | Traditional Search Engine | Metasearch Engine |
|---|---|---|
| Database | They compile a physical database or catalogue of the web | They create the virtual database |
| Coverage | They can search the surface web and to some extent deep web | They can search the deep web |
| Precision | Less precision of result | More precision than search engine by using query modification |
| Result Relevancy | The search engines might give irrelevant results | They used clustering algorithm to provide relevant result |
| Response Time | Search Engine takes fraction of second to give result | It takes approximately twice the time with respect to search engines |
| Relevant result time | User has to spend more time to get relevant result | It presents relevant result in less time by querying to various search engine |
| Network Bandwidth | It consumes lot of network bandwidth for continuing updation of database | Consumes little bandwidth as it does not have to update their database |
| Dependency | Search Engines depend upon their ranking algorithm and their database | Metasearch engines depend upon the search engine to be used |
| Redundancy | Redundancy of same web page in the result | They omit out the duplicate entry |
| Hardware Requirement | Requires large investment in hardware for storage of database | Requires lesser investment in hardware |
| Implementation cost | Involves high initial implementation cost and high maintenance cost | Relatively low implementation and maintenance cost |

Table 2. Comparative Study of Traditional Search Engines and Metasearch Engines

## 5. CONCLUSIONS AND FUTURE WORK

In this work, we have compared search engines and metasearch engines on the basis of various parameters and found that metasearch engines are useful if the user is looking for a unique term or phrase or the user simply wants to run a couple of keywords. The study based on survey of articles and examination of test experiments for the quality of the metasearch engines, it has been found that metasearch engines have proved to be more effective than conventional search engine. The following four reasons have been concluded that the metasearch engine be preferred over the traditional search engines. 1) It is better to query metasearch engines to obtain the most relevant result. 2) Meta searching is an excellent approach for both broad and shallow searches. 3) For keywords of an unfamiliar subject, the better way to discover search terms is to see how they appear in a cross section of documents across the web. 4) Metasearch engine is an excellent way to know about the different search engines, their strengths, weaknesses, and types of queries they can handle.

Our future work will be to explore what kinds of metasearch engines will be required to meet the needs of the future. One interesting development in this direction has already taken place in the form of meaning-based search engine.

The metasearch engine's functionality is crippled by the slow response rate. In order to make then usable, we need them to run much faster. Currently, the metasearch engines allow the user to define the priority of the search engines. It would be interesting to determine the priority based on statistical information or learning-based methods. We are also working on to design a metasearch engine based on mobile agents.